\documentclass[aps,prl,twocolumn,showpacs]{revtex4-1}
\usepackage{graphicx}
\usepackage{subfigure}
\usepackage{amsmath}
\usepackage{epsfig}
\usepackage{epstopdf}

\DeclareMathOperator{\sgn}{sgn}

\begin{document} 
\title{Mean-potential law in evolutionary games} 
\author{Pawe\l \; Na\l\c{e}cz-Jawecki}
\affiliation{College of Individual Studies in Mathematics and Natural Sciences, University of Warsaw,  ul. Banacha 2C, 02-097 Warsaw, Poland, email:pawel.nalecz-jawecki@student.uw.edu.pl}
\author{Jacek Mi\c{e}kisz}
\affiliation{Institute of Applied Mathematics and Mechanics, University of Warsaw,  ul. Banacha 2,  02-097 Warsaw, Poland, email: miekisz@mimuw.edu.pl}

\begin{abstract}
The Letter presents a novel way to connect random walks, stochastic differential equations, and evolutionary game theory. 
We introduce a new concept of potential function for discrete-space stochastic systems. It is based on a correspondence between one-dimensional 
stochastic differential equations and random walks, which may be exact not only in the continuous limit but also in finite-state spaces. 
Our method is useful for computation of fixation probabilities in discrete stochastic dynamical systems with two absorbing states. 
We apply it to evolutionary games, formulating two simple and intuitive criteria for evolutionary stability of pure Nash equilibria 
in finite populations. In particular, we show that the $1/3$ law of evolutionary games, introduced by Nowak et al 
[Nature, 2004], follows from a more general mean-potential law. 
\end{abstract}

\maketitle

\vspace{3mm}

{\bf Introduction.} Many biological and social processes can be modeled within the framework 
of evolutionary game theory \cite{maynard1,maynard2,nowakbook}. Players with different behaviors (strategies) interact with each other and receive payoffs; the higher the average payoff, the greater
the chance that the player's offspring will replace another player, and so the population changes its composition over time -- in favour of more profitable strategies. Maynard Smith and Price \cite{maynard1,maynard2} 
introduced the fundamental notion of evolutionarily stable strategy (ESS). If everybody plays such a strategy, and a small group of mutants 
playing a different one arrives, then the mutants will be eliminated from the population due to natural selection. The dynamical interpretation of evolutionarily stable strategies was later provided 
by several authors \cite{ty,hof,zee}. They proposed a system of differential equations, called replicator equations, 
which describe how the abundance of particular strategies in the population changes in time. It is known that any evolutionarily stable strategy 
is an asymptotically stable stationary point of such a dynamics \cite{hofbook}. 

Replicator equations describe the population behavior 
in the limit of an infinite number of individuals. However, real populations are finite. Stochastic effects connected 
with random matching of players, mistakes in decisions, and biological mutations can play a significant role in such systems. 
Therefore to describe real systems one should deal with stochastic dynamics of finitely many interacting individuals.
Hence the concept of evolutionarily stable strategy was reformulated by Nowak et al. \cite{nowak} so that it takes into account 
 the stochastic fluctuations in finite populations as well.

In this paper we discuss a simple stochastic evolutionary dynamics that may be described as a (biased) 1D random walk with two absorbing states. 
We show a way to visualize the dynamics, compute fixation probabilities, and easily decide whether a strategy is evolutionarily stable or not.

{\bf Pairwise comparison stochastic dynamics.} Let us consider a population of $N$ players who have two strategies at their disposal: $A$ and $B$. 
The individuals take part in a two-player symmetric game given by the following payoff matrix:  
\begin {align}
\begin{matrix}
  &\vline& A & B \\ \hline
A &\vline& a & b \\
B &\vline& c & d \\
\end{matrix}
\end{align} 
where the $ij$th entry, $i,j = A, B$, is the payoff of a player using the strategy $i$ against a player with the strategy $j$.
Throughout the paper we assume that $a>c$ and $d>b$, so that 
$A$ and $B$ are both pure Nash strategies (i.e. $(A,A)$ and $(B,B)$ are symmetric Nash equilibria) and $x^* = (d-b)/(d-c+a-b)$ 
is an unstable mixed Nash equilibrium. An example of this setup is the Stag-hunt game \cite{maynard2,nowakbook}.

Let us denote by $x$ the fraction of individuals playing $A$. As it is usually done in the context of population dynamics based on game contests, 
we assume that players receive payoffs  equal to their expected rewards from a game against a random opponent. 
Hence the payoffs of $A$ and $B$ are given by 

\begin{subequations}
\begin{align}
f_{A}(x) = ax + b(1-x) \\
f_{B}(x) = cx + d(1-x)
\end{align}
\end{subequations}

The standard pairwise comparison stochastic dynamics is as follows \cite{claussen}. At any discrete moment of time, two players are chosen at random, 
a focal one (say an $X$-strategist, $X\in\{A,B\}$) and a model one (say a $Y$-strategist). The focal player adopts the strategy of the model one 
with probability depending on the payoff differences, namely

\begin{equation}
p(X \rightarrow Y) = \frac{1}{2}+\frac{1}{2}\cdot g_w(f_{Y} - f_{X})
\end{equation}
where $g_w$ is an arbitrary antisymmetric, non-decreasing function and $w$ is a parameter called selection pressure such that if $w\rightarrow 0$, then $g_w\rightarrow 0$.
Particular choices of $g_w(\Delta f)$ define various dynamics. 
Common possibilities include $g_w =w \sgn \Delta f$ (binary model; ``stronger-wins''), $g_w =w(\Delta f)/(\Delta f_{\mathrm{max}})$ (`linear advantage') 
and $g_w =(1-e^{-w\Delta f})/(1+e^{-w\Delta f})$ (Fermi rule). 

The replicator equation ${dx/dt=x(1-x)\cdot g_w(f_A(x)-f_B(x))}$ 
describes time evolution of $x$ in the infinite population -- the frequency of the more profitable strategy 
increases at rate proportional to $g_w$ and to the frequency of encounters between $A$ and $B$ \cite{hofbook}. 
Under our assumptions, $x=0$ and $x=1$ are two asymptotically stable stationary points and $x^*$ is an unstable one. 

The discrete model can be described by a one-dimensional random walk -- a Markov chain with $(N+1)$ states: $\{0, \frac{1}{N},\dots,1\}$, 
of which two, $0$ and $1$, are absorbing, and with transition probabilities 
\begin{subequations}
\label{eq_markovchain}
\begin{align} 
p(x \rightarrow x + \frac{1}{N}) = x (1-x) p(B\rightarrow A) =: a_+(x) \\
p(x \rightarrow x - \frac{1}{N}) = x (1-x) p(A\rightarrow B) =: a_-(x)
\end{align}
\end{subequations}

{\bf Evolutionary stability.} According to the classical definition for infinite populations, strategy $B$ is considered {\bf evolutionarily stable} (ESS) 
if in a state with a sufficiently small fraction of $A$-players, the expected payoff of $B$ 
is greater than that of $A$. Hence a uniform population of $B$-players is protected by selection against invasion 
of a small number of mutants playing $A$ \cite{maynard1}. In games with two pure Nash equilibria, 
both $A$ and $B$ strategies are evolutionarily stable. 

In finite populations, the classical concept was extended to take into account the possibility 
 of a single mutant taking over the whole population due to stochastic fluctuations. 
 The following definition was proposed in \cite{nowak}.

Strategy $B$ is {\bf evolutionarily stable in finite populations oz size $N$} []ESS($N$)] if in the situation when $(N-1)$ individuals  play $B$, and a single one plays $A$,  two inequalities hold: (i) the $A$-player has a smaller expected payoff than a $B$-player (resilience to invasion)
and (ii) the probability of fixation of strategy $A$ and extinction of $B$ is smaller than $1/N$ (resilience to replacement). 

The first inequality is a discrete analogue of the ESS condition for infinite populations. 
The second one tells us whether the chance for the population to acquire the invador's strategy 
(and the dominating strategy to go extinct) is smaller than  
in the neutral-selection model (i.e. the unbiased random walk).

To check if $B$ is an ESS($N$) one needs to compute the fixation probability of $A$, supposing we start from a population consisting 
of one $A$-player and $(N-1)$ $B$-players. In other words, we ask what is the probability of a biased random walk starting at position $\frac{1}{N}$
to reach $1$ without visiting $0$.  

Our novel approach is as follows. We start with a one-dimensional stochastic process 
on the continuous space $[0,1]$. We use a continuous-space potential function to get formulas for fixation probabilities. 
Then we construct a discrete-space process corresponding to the continuous one and show how reversal of the procedure 
lets us compute fixation probabilities in the discrete case and decide quickly whether a strategy is evolutionary stable in finite populations.

{\bf Potential in continuous and discrete processes.} Let us consider a process $X_t$ with continuous time $t$ 
solving the It\^{o} stochastic differential equation
\begin{equation}
\label{eq_SDE_potential}
dX_t=-\frac{1}{2}\varphi'(X_t)dt+dW_t
\end{equation}
where $\varphi' = d\varphi/dx$ is a derivative of an arbitrary (smooth enough) function $\varphi$, defined 
on the interval $[0,1]$. We will call $\varphi$ the {\it potential function}.  The above process can be seen as a movement 
of a point down the slope of $\varphi$, 
disturbed by the Brownian diffusion of unit variance. It was introduced in game theory models in \cite{peyton}.
Let us set absorbing boundary conditions at $x=0$ and $x=1$, and denote by  $\rho(x_0)$ the probability of fixation at point $x=1$ 
provided that we start from $x=x_0$. This probability satisfies the stationary Kolmogorov backward equation \cite{vankampen,grimmett},
\begin{equation}\label{kolmoback}
\frac{1}{2}\varphi ' \rho' - \frac{1}{2} \rho'' = 0
\end{equation}
which combined with boundary conditions $\rho(0)=0$ and $\rho(1)=1$ gives
\begin{equation}
\label{eq_rho}
\rho(x_0)= \frac{\int_{0}^{x_0}e^{\varphi(x)} dx}{\int_{0}^{1}e^{\varphi(x)}dx}
\end{equation}

Let us note that simmilar diffusion approximations were used to calculate fixation probabilities in \cite{ohtsuki,traulsen,cressman}.
However, it is possible to contruct a discrete process corresponding {\it exactly} to continuous one defined above. Consider two objects, a \emph{ball} and a \emph{pawn}, 
located initially at the same point $x_0\in \{0,\frac{1}{N}, \dots, 1\}$. 
Let the ball move according to the continuous process (\ref{eq_SDE_potential}). 
Whenever the ball reaches any of the lattice points $\{0,\frac{1}{N}, \dots, 1\}$, we move the pawn to the ball. 
It follows from Eq.~(\ref{eq_rho}) that the motion of the pawn is a discrete process with transition probabilities given by 

\begin{subequations}
\label{eq_p}
\begin{align}
a_+(x_0) = \frac{\int_{x_0-1/N}^{x_0}e^{\varphi(x)} dx}{\int_{x_0-1/N}^{x_0+1/N}e^{\varphi(x)}dx}\\
a_-(x_0)= \frac{\int_{x_0}^{x_0+1/N}e^{\varphi(x)} dx}{\int_{x_0-1/N}^{x_0+1/N}e^{\varphi(x)}dx}
\end{align}
\end{subequations}

Note that the pawn and the ball will reach an absorbing state ($0$ or $1$) exactly at the same time. 
Therefore the probability of the pawn's fixation at $1$, provided it started at $x_0$, may be computed in the same way 
as for the ball -- with Eq.~(\ref{eq_rho}). Various choices for $\varphi$ give us different discrete dynamics -- setting $\varphi$ 
appropriately we can recover any desired discrete random walk. 
As long as the random walk is one-dimensional (i.e. there are only two possible strategies), the potential function always exists, though it is not unique. 
To get one, it is possible to assume that $\varphi$ is linear on every interval $[\frac{k}{N}, \frac{k+1}{N}]$, 
set values at $x=0$ and $x=\frac{1}{N}$ arbitrarily and recurrently find the values at points $\frac{2}{N}, \dots, 1$ by solving Eqs.~(\ref{eq_p}).
This was done to obtain the exact potentials shown in Fig.~1.

There is a straightforward connection between the potential function and the evolutionary stability.
Indeed, one can easily observe that a strategy is an ESS (in infinite populations) iff the potential has a local minimum 
at the boundary point corresponding to that strategy. In other words, strategy $B$ is evolutionarily stable 
iff $\varphi$ is increasing at $x=0$, whereas $A$ is stable whenever the potential decreases at $x=1$.
This argument holds also for condition (i) of ESS($N$) in discrete-space systems, apart from that one should look at finite differences 
rather than derivatives when determining monotonicity at $x=0$ and $x=1$. Note that in games with two pure Nash equilibria,
condition (i) is always satisfied both for $A$ and $B$.

As for condition (ii), we now present two general laws which are helpful with calculating fixation probabilities.
The first one is just the reformulation of condition (ii) in terms of the potential function.
\vspace{1mm}

{\bf Mean-exponential-potential law.} 
The following inequality is equivalent to condition (ii):
\begin{equation}
\label{eq_MEPP_precise}
\frac{1}{1/N} \int_{0}^{1/N}e^{\varphi(x)}dx < \int_{0}^{1}e^{\varphi(x)}dx
\end{equation}
This is an immediate consequence of Eq.~(\ref{eq_rho}). 
It can be called the mean-exponential-potential law: strategy $B$ satisfies condition (ii) of ESS($N$) iff the average of $e^{\varphi(x)}$
is smaller on $[0,\frac{1}{N}]$ than on the whole space $[0,1]$.
In the limit $N \rightarrow \infty$  we get

\begin{equation}
\label{eq_MEPP}
e^{\varphi(0)} < \int_{0}^{1}e^{\varphi(x)}dx
\end{equation}

{\bf Mean-potential law.} In the weak-selection case, that is when $w \ll \frac{1}{N}$, we can assume that $\varphi(x) = \varphi(0)$ 
plus some small $x$-dependent perturbation. Consequently Eq.~(\ref{eq_MEPP_precise}) can be approximated by a linear term 
in the expansion of $e^{\varphi(x)}$, giving that in the weak-selection case, $B$ is an ESS($N$) iff 

\begin{equation}
\label{eq_MEP_precise}
\frac{1}{1/N}\int_{0}^{1/N}\varphi(x)dx <\int_{0}^{1}\varphi(x)dx
\end{equation}
Again, in the limit $N \rightarrow \infty$ we get
\begin{equation}
\label{eq_MEP}
\varphi(0) <\int_{0}^{1}\varphi(x)dx
\end{equation}

This is an easy-in-use criterion, and we show in the following paragraphs how to apply it to two classical pairwise comparison models 
in order to get conditions for evolutionary stability of given strategies.

{\bf Model 1: Stronger wins.} In such a dynamics, the probability of adopting the opponent's payoff depends only 
on the sign of the payoff difference -- the transition probabilities are

\begin{align}
a_\pm (x)=\frac{1}{2}\pm\frac{w}{2}\cdot \sgn(x-x^*) 
\end{align}
Here and further on we exclude steps in which the state of the population does not change; 
that happens in particular when the focal and model players play the same strategy. 
Hence the term $x(1-x)$ in Eq.~(\ref{eq_markovchain}) is dropped.

As the drift is a step function, the potential $\varphi$ is expected to be proportional to $-|x-x^*|$. 
Indeed, one may check the following potential satisfies Eqs.~(\ref{eq_p}):

\begin{equation}
\varphi(x)=-\left(N\ln \kappa \right) |x-x^*|
\end{equation}
with $\kappa = \frac{1+w}{1-w}$. Therefore one may use Eq.~(\ref{eq_rho}) to calculate the probability that one $A$-player 
overtakes the whole population. We get

\begin{equation}\label{fixingprob}
\rho(1/N) = \frac{\kappa^{-(x^*-1/N)N} - \kappa^{-x^*N}}
{2-\left(\kappa^{-x^*N}+\kappa^{-(1-x^*)N}\right)}
\end{equation}

Now, condition (ii) of ESS($N$), or equivalently the mean-exponential-potential law (\ref{eq_MEPP_precise}), gives us
\begin{equation}
\label{eq_xkr_sgn}
x^* >  1+\frac{\ln \left(1-\sqrt{1-\kappa^{-N}(N(\kappa-1)+1)}\right)}{N\ln \kappa} =: x_{cr}
\end{equation}

This is an exact condition on the mixed Nash equilibrium $x^*$ for strategy $B$ to be evolutionarily stable.
By symmetry, $A$ is evolutionarily stable if $x^* < 1 - x_{cr}$. Of course for $x_{cr} < x^* < 1-x_{cr}$, 
both strategies are evolutionarily stable.
 
In the weak-selection case ($w \ll 1/N$), we compute the limit of the right-hand side of (\ref{eq_xkr_sgn})
as $w N \rightarrow 0$ and we get
\begin{equation} 
x^* > 1-\sqrt{\frac{1}{2}-\frac{1}{2N}}
\end{equation}

Finally we take $N\rightarrow \infty$ and obtain
\begin{equation}\label{newlaw} 
x^* > 1-\sqrt{\frac{1}{2}}
\end{equation}

By analogy with $1/3$ law \cite{nowak}, this could be called the law of $\left(1-\sqrt{\frac{1}{2}}\right)$. Inequality (\ref{newlaw}) 
can be also easily obtained from the mean-potential law as the mean potential on $[0,1]$ is equal to $-\frac{c}{2}((x^*)^2+(1-x^*)^2)$ 
for ${c = N\ln \kappa }$ and $\varphi(0) = -cx^*$.

{\bf Model 2: Linear advantage.} In this model \cite{claussen}, the probability of adopting the model player's strategy depends linearly
on the difference of payoffs:

\begin{align}
a_\pm (x)=\frac{1}{2}\pm\frac{w}{2}\cdot \frac{f_A-f_B}{\Delta f_{\mathrm{max}}} 
\end{align}
where $\Delta f_{\mathrm{max}} = \max \left\{ \left| a-c \right| ,\left| d-b\right| \right\}$ is the maximal possible payoff difference. 
Setting for simplicity of notation $k=2\cdot\frac{a-b-c+d}{\Delta f_{\mathrm{max}}}$ we can write

\begin{align}
\label{eq_a_liniowy}
a_\pm (x)=\frac{1}{2}\pm\frac{wk}{4}(x-x^*) 
\end{align}
One can notice that the expected value of one-step displacement is $\frac{1}{N}(a_+-a_-)=\frac{wk}{2N}(x-x^*)$, 
and its variance is $\frac{4a_+a_-}{N^2}=\frac{1}{N^2} + O(\frac{w^2k^2}{N^2})$. 
From Eq.~(\ref{eq_SDE_potential}) one gets that in the continuous case $\varphi' = -2\frac{\mu}{\sigma^2}$. Hence the natural candidate for the potential function is 

\begin{equation}
\tilde\varphi=-2N\int \frac{wk}{2}(x-x^*) =-\frac{wkN}{2}(x-x^*)^2
\end{equation}

Unfortunately such a function does not satisfy Eqs.~(\ref{eq_p}). If it did, the conditions for evolutionary stability 
would follow immediately: the mean value of that potential on $[0,1]$ would be 
$-\frac{c}{6}\left((1-x^*)^3~-~(-x^*)^3\right)=-\frac{c}{2}\left((x^*)^2-x^*+\frac{1}{3}\right)$ for $c=wkN$,  
and as $\tilde\varphi(0)=-\frac{c}{6} (x^*)^3$ one would get from Eq.~(\ref{eq_MEP}) that $\rho(1/N) < 1/N$ iff $x^*>\frac{1}{3}$. Note that this is the Nowak's $1/3$-law \cite{nowak}. 

By this calculation we proved that the law holds for a discrete stochastic process for which $\tilde\varphi$ is the real potential. 
However, one can show that fixation probabilities in such an auxiliary model and the true one are close to each other, 
at least for large $N$ and weak selection $w$. Intuitively, this is because the difference in transition probabilities 
between the true and the auxiliary model are of order $\frac{1}{N^3} |\tilde\varphi''| \sim \frac{wk}{N^2}$,
and fixation is reached on average in $O(N+\frac{N^2}{(wkN)^2+1})$ steps, 
which means that the expected number of steps in which the models disagree decreases to zero
as $wk\cdot (\frac{1}{N}+ \frac{1}{(wkN)^2+1})$. However, a precise proof of this fact is beyond the scope
of this Letter.

\begin{figure}
\label{fig}
\includegraphics[width=0.5\textwidth]{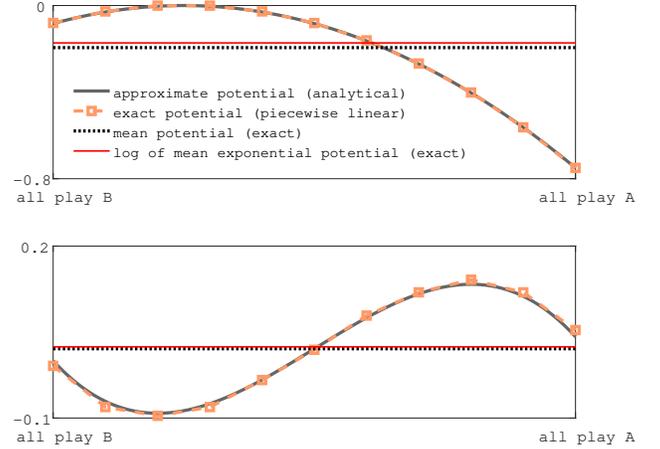}
\caption{
Potential function, guessed from Eq.~(\ref{eq_varphi_and_a}) (approximate) 
and computed numerically (exact), for two models with linear advantage update: 
(top) a stug-hunt game with $x^{*}=0.25$, 
(bottom) a three-player game with the following payoffs: 
$A$ receives $4$ when playing against $AA$ or $BB$ and $0$ otherwise, 
$B$ gets $17$ against $AB$ or $BA$, otherwise nothing.
Number of players and selection pressure are $N=10$ and $w=0.1$. 
In the first model, both strategies are pure Nash equilibria. 
$A$ is evolutionarily stable in finite populations
because at the point when all play $A$ the potential (i) has a local minimum
and (ii) is below the mean potential. 
In the second model, both pure strategies are unstable -- $A$ does not satisfy condition (ii), 
whereas $B$ fails to meet condition (i).
}
\end{figure}

{\bf Discussion.} We introduced a concept of potential function for discrete random walks. 
The function is defined on the whole interval $[0,1]$, which is in contrast to previous approaches \cite{mondererShapley}, 
where the potential was defined on a discrete set of points. This allowed us to get an immediate correspondence 
with continuous-space systems and stochastic differential equations. As mentioned before, the potential exists for any game with two strategies. 

From the numerical point of view, 
this procedure is by no means easier than direct computation of fixation probabilities (see, e.g., \cite{nowak} for the exact formula). 
However, one may usually guess or compute an approximate potential 
by comparing the expected displacement in one step and its variance with Eq.~(\ref{eq_SDE_potential}):
\begin{equation}
\label{eq_varphi_and_a}
\varphi' \approx - 2\frac{\mu}{\sigma^2}= -2N\frac{a_+-a_-}{4a_+a_-} \approx -2Ng_w
\end{equation}
This approximation is likely to give good estimates for fixation probabilities 
and lets us easily assess evolutionarily stability of a given strategy with the mean-potential law.
If necessary, a quality check of the approximation is possible, both numerical (as shown in Fig.~1) and analytical (as sketched in Model 2).

{\bf Other directions.} The potential function approach is more general than presented here. 
Mean-potential laws can be applied to {\bf multi-player games} \cite{broom,jabuk}. In Fig.~1 
we present an example of a three-player game with a unique pure ESS 
which is not an ESS($N$). This is caused by the presence of an asymptotically stable interior point 
(an evolutionarily stable mixed Nash equilibrium). In two-player games with evolutionarily stable strategies, 
at least one of them is evolutionarily stable in finite populations. 

As an alternative to the ESS($N$), one may consider the concept of {\bf stochastic stability} \cite{freidlin} introduced in evolutionary games in \cite{peyton} 
and analyzed, e.g., in \cite{kmr,peyton2,redondo,redondobook,cime}. Models investigated in those papers allow spontaneous mutations, 
occurring with the rate $\epsilon = 1 - w$. That means that even in homogeneous populations, individuals can change their strategy.
The two absorbing states disappear and we obtain an ergodic Markov chain with a unique stationary distribution. 
A population state $x_0$ is called {\it stochastically stable} if in the limit $\epsilon \rightarrow 0$ the stationary distribution 
is concentrated in $x_0$. The main result of \cite{kmr} is that in the ``stronger-wins'' model, the strategy $B$ (i.e. the state $x=0$) 
is stochastically stable iff it has a bigger basin of attraction, that is if $x^* > 1/2$. 
This can be compared with the limit $w \rightarrow 1$ of (\ref{eq_xkr_sgn}). In this strong-selection case $x_{cr}=1$, 
and we get that both pure strategies are evolutionarily stable.
However, the potential function allows us to formulate a more general statement: 
if there exists a potential function $\varphi_\epsilon$ such that $\varphi_\epsilon(x)\rightarrow +\infty$ with $\epsilon\rightarrow 0$ 
for all $x\neq x_0$ and $\varphi_\epsilon(x_0)\rightarrow 0$, then under some convergence conditions the state $x_0$ is stochastically stable. 

{\bf Summary.} We presented a novel way to analyze finite-state one-dimensional random walks.
It is based on an exact transformation of continuous stochastic processes into discrete ones which preserves fixation probabilities.
It allows us to transfer the intuitive concept of potential function from the continuous onto the discrete case.
Using the potential function, we formulated two simple criteria to determine whether the probability of fixation in a given boundary point is higher compared to an unbiased random walk.
We called them the mean-exponential-potential law (Eq.~(\ref{eq_MEPP})) and the mean potential law (Eq.~(\ref{eq_MEP})).
Both of them have immediate implications for evolutionary game theory, allowing quick identification of evolutionarily stable strategies in finite populations. 
We illustrated their usage on two examples of 2-player evolutionary games -- in particular, we re-derived the 1/3 law, which occurred to be a special case of the mean-potential law 
-- and suggested a few other applications. The presented ideas are more general though, and we  hope they will be useful in various fields of study.

\vspace{2mm}

\begin{acknowledgments} 
JM would like to thank National Science Centre (Poland) for a financial support under the grant 2015/17/B/ST1/00693.
\end{acknowledgments}

\end{document}